\documentclass[%
reprint,
superscriptaddress,
amsmath,amssymb,
aps,
prd,
]{revtex4-1}
\usepackage{tikz}
\usepackage{graphicx}% Include figure files
\usepackage{dcolumn}% Align table columns on decimal point
\usepackage{csquotes}
\usepackage{braket}
\usepackage{bm}% bold math
\usepackage{amsmath}

\begin{document}

\preprint{APS/123-QED}

\title{ A Search for deviations from the inverse square law of gravity at nm range using a pulsed neutron beam}

\author{Christopher C. Haddock}
\affiliation{Nagoya University, Furocho, Chikusa Ward,
Nagoya, Aichi Prefecture 464-0814, Japan}

\author{Noriko Oi}
\affiliation{Nagoya University, Furocho, Chikusa Ward,
Nagoya, Aichi Prefecture 464-0814, Japan}

\author{Katsuya Hirota}
\affiliation{Nagoya University, Furocho, Chikusa Ward,
Nagoya, Aichi Prefecture 464-0814, Japan}

\author{Takashi Ino}
\affiliation{High Energy Accelerator Research Organization KEK 1-1 Oho, Tsukuba, Ibaraki, Japan, 305-0801}

\author{Masaaki Kitaguchi}
\affiliation{Nagoya University, Furocho, Chikusa Ward,
Nagoya, Aichi Prefecture 464-0814, Japan}

\author{Satoru Matsumoto}
\affiliation{Department of Physics, Kyushu University
  744 Motooka, Nishi-ku, Fukuoka, Japan}

\author{Kenji Mishima}
\affiliation{High Energy Accelerator Research Organization KEK 1-1 Oho, Tsukuba, Ibaraki, Japan, 305-0801}

\author{Tatsushi Shima}
\affiliation{Research Center for Nuclear Physics, Osaka University
 10-1 Mihogaoka, Ibaraki, Osaka, 567-0047} 

\author{Hirohiko M. Shimizu}
\affiliation{Nagoya University, Furocho, Chikusa Ward,
Nagoya, Aichi Prefecture 464-0814, Japan}
  
\author{W. Michael Snow}
\affiliation{Department of Physics, Indiana University
727 E. Third St., Swain Hall West, Room 117, Bloomington, IN 47405-7105}
 
\author{Tamaki Yoshioka}
\affiliation{Research Center for Advanced Particle Physics, Kyushu University
744 Motooka, Nishi-ku, Fukuoka, Japan}

\begin{abstract} %%%%%%%%%%%%%%%%%%%%%%%%%% Abstract %%%%%%%%%%%%%%%%%%
We describe an experimental search for deviations from the inverse square law of gravity at the nanometer length scale using neutron scattering from noble gases on a pulsed slow neutron beamline. By measuring the neutron momentum transfer ($q$) dependence of the differential cross section for xenon and helium and comparing to their well-known analytical forms, we place an upper bound on the strength of a new interaction as a function of interaction length $\lambda$ which improves upon previous results in the region $\lambda < 0.1\,$nm, and remains competitive in the larger $\lambda$ region. A pseudoexperimental simulation developed for this experiment and its role in the data analysis described. We conclude with plans for improving sensitivity in the larger $\lambda$ region. 
\end{abstract}

\maketitle

%\tableofcontents

\section{Introduction} %%%%%%%%%%%%%%%% Introduction %%%%%%%%%%%%%%%%%%
\label{sec:intro}

%100 introduction

It is known that there are four fundamental interactions in Nature: the strong, weak, electromagnetic, and gravitational forces. The first three are unified within the framework of a renormalizable relativistic quantum gauge theory known as the Standard Model (SM) of particle physics. Gravity is described separately using the theory of General Relativity (GR). GR is fully consistent with all experiments and observations to date, including the recent dramatic discovery of gravitational waves\,\citep{Abbott2016}. However this classical theory of gravity cannot be made consistent with quantum mechanics in a straightforward way as has been done successfully for the other interactions. Although the mathematical success of string theory shows that a theory of gravity which is consistent with quantum mechanics and the other known interactions is possible, there are no direct experimental tests of this theory of gravity so far. Experiments which probe gravity in new regimes are therefore of fundamental interest.   

The force of gravity is confirmed to follow an inverse-square law (ISL) famously known as \enquote{Newton's Law of Universal Gravitation}. This law can be derived within GR in the nonrelativistic limit in a weak gravitational field. Its validity has been verified experimentally down to distances of less than $1$\,mm\,\citep{Adelberger09}. In a quantum theory it would be equivalent to the statement that the mass of the graviton which transmits the interaction is zero. The recent observation of a neutron star merger in both gravitational waves and throughout the electromagnetic spectrum using multimessenger astronomy\,\citep{Abbott2017} places a very stringent limit on the difference between the speed of light and the speed of gravitational waves and therefore a stringent upper bound on the mass of the graviton. The long-distance component of the gravitational interaction is therefore very well understood. 

From a strictly experimental point of view, however, the nature of gravity at short ranges is very poorly constrained. Many alternative theories of gravity possess an additional component of the gravitational interactions which involve the exchange of massive quanta and therefore only extends to short ranges. Other theories which try to explain why gravity is so weak compared to the other interactions of Nature also produce short-distance modifications to the theory. At the end of 1990 an interesting model, namely the Large Extra Dimension theory, was advocated by N. Arkani Hamed et al.\,\citep{Arkani98, Adelberger03}, which suggests that gravity is compactified within extra spatial dimensions of finite extent. One very interesting prediction of this theory was that, if one assumes that the strength of the gravitational interaction is the same as that of other forces at the electroweak scale to solve the so-called \enquote{hierarchy} problem of particle theory, then one should see  deviations from the ISL at distances below $1$\,mm if the number of extra spacetime dimensions is 2. Other assumptions within this general framework lead to different distance regimes where one might observe ISL violations. 

One can also interpret searches for ISL violations as constraints on possible new interactions of nongravitational origin. The very wide variety of speculations which can lead to such interactions has been reviewed recently\,\citep{Safronova2017}. For weakly-coupled interactions between nucleons and electrons in the nonrelativistic limit, a number of parametrizations for such interactions have been developed which are largely model-independent under the assumption that the new interaction is local\,\citep{Dobrescu06, Brax17}. These parametrizations typically produce ISL violations with a Yukawa-like exponential falloff with distance multiplied by some power of $1/r$.  

Many experiments have been carried out to search for a deviation from the ISL at short distances.  Torsion balances\,\citep{Kapner2007} have been used to search for ISL deviations at the $50$ micron scale and above. Data used to measure Casimir interactions have been first reanalyzed and later the relevant apparatus modified specifically to search for ISL deviations on micron scales\,\citep{Mohideen98, Chiaverini03, Sushkov11, Chen16}. Experiments using laser-levitated microspheres to search in this distance regime have been proposed or are in progress\,\citep{Geraci10, Geraci15}. At these distance scales interatomic forces generated by the electric polarizability of atoms contribute a serious experimental background.  

Experiments using neutrons can be used to search for ISL deviations at submicron scales and, due to their net electrical neutrality, are not constrained by large electromagnetic background effects.  %search for possible new net electrican extremely  electric, which for neutrons is determined by the strong forces between quarks rather than electromagnetic forces
The ultimate limit in sensitivity using neutrons set by the present brightness of free neutron sources has not yet been reached and so there is room for discovery. Nesvizhevsky et al.\,\citep{Nez08} reanalyzed many neutron measurements originally conducted for other purposes to constrain such interactions. In particular they reanalyzed an experiment designed to measure the low energy neutron-electron scattering amplitude\,\citep{Krohn73} which measured a forward/backward asymmetry in the angular distribution of neutron scattering on noble gases.   

In this paper we describe a dedicated experiment which probes the ISL at the nanometer length scale using neutron-noble gas scattering. Our experiment is similar to one recently conducted by Kamiya {\it et al.} using a small angle neutron scattering instrument at the HANARO research reactor in Korea\,\citep{kamiya}. Ours however is the first experiment of its kind to utilize the time-of-flight analysis capabilities of a pulsed spallation neutron source, allowing us to probe a relatively wide range of neutron momentum transfer.

\section{Methodology} %%%%%%%%%%%%%%%% Methodology %%%%%%%%%%%%%%%%%%
\label{sec:method}

For the rest of this paper we assume that this unknown interaction comes from a single exchange of a massive spinless boson. The nonrelativistic limit of a quantum field theory describing the exchange of a single massive boson generates a Yukawa potential in position space. We treat this interaction as a perturbation to the Newtonian force so that the potential $V(r)$ between masses of $m$ and $M$ can be described as
\begin{equation}
V(r) = -G_N \frac{mM}{r}(1+\alpha\,e^{-r/\lambda})
\label{eq:vofr}
\end{equation}
where $r$ is distance between the masses, $\alpha$ parametrizes the strength of the short-range interaction relative to gravity, $\lambda$  is the Compton wavelength of the exchange boson, and $G_N$ is the gravitational constant.

In order to understand how a neutron-noble gas scattering experiment can be made sensitive to the Yukawa term in Eq.\,\ref{eq:vofr}, we look at its contribution to the total coherent neutron-atom scattering amplitude. The total neutron-atom scattering amplitude for the case of unpolarized neutrons can be written in terms of the momentum transferred from the neutron to the scattering center, $q$, as\,\citep{sears}

\begin{equation}
b(q) = b_\text{N}+b_\text{E}(q)+b_\text{M}(q)+b_\text{Y}(q) 
\label{eq:btot}
\end{equation}

\noindent where $b_\text{N}$ is the $q$-independent low energy s-wave nuclear scattering amplitude from the strong interaction, $b_{\text{E}}(q)$ describes interactions between the neutron's charge distribution and the atomic electric field, $b_\text{M}(q)$ arises from interactions between the neutron's magnetic dipole moment and the slowly varying electric and magnetic fields of the scattering centers, and $b_\text{Y}(q)$ describes the contribution from the possible exotic Yukawa-like interaction. For the case of diamagnetic atoms (such as the noble gases) with very low incoherent scattering cross sections the contribution from $b_\text{M}(q)$ to the differential cross section is at least $10^{-3}$ times smaller than the other nongravitational terms, and we therefore neglect its contribution from here on.

Each interaction generates a scattering amplitude with a distinct $q$-dependence. The electric scattering amplitude is written as 

\begin{equation}
b_\text{E}(q) = -b_\text{e} Z\bigg(1-\frac{1}{1+(\frac{q}{q_0})^2}\bigg)
\label{eq:be}
\end{equation}
% b_e = (-0.00129\pm 0.00003)
where $b_\text{e} = -1.32(4)\times\,10^{-3}$\,fm\,\citep{KWK,RussiaNE} is the neutron-electron scattering amplitude, $Z$ is the atomic charge number, and $q_0$ is a parameter which quantifies the form factor of the electron spatial distribution of the atom. For the noble gases, $q_0\sim(30-70)$\,nm$^{-1}$\citep{sears}. The scattering length $b_\text{Y}(q)$ is computed simply in the first Born approximation as the Fourier transformation of Eq.\,\ref{eq:vofr} into the $q$ plane and is given as 

\begin{equation}
b_\text{Y}(q) = \alpha \bigg(\frac{2G_\text{N}m_\text{n}^2M_\text{A}}{\hbar^2}\bigg)\frac{1}{\lambda^{-2}+q^2}
\label{eq:by}
\end{equation}

\noindent where $M_\text{A}$ is the mass of the atom, $m_n$ is the neutron mass, and $\hbar$ is the reduced Planck's constant. Comparing $b_\text{E}(q)$ to $b_\text{Y}(q)$ we see that for values of $\lambda \sim $\,nm, $b_\text{E}(q)$ is dominant in the higher $q$ region while $b_\text{Y}(q)$ is dominant in the lower $q$ region. The existing experimental limits on the value of $b_\text{Y}(q)$ are of order $10^{-3}b_\text{N}$. The square of the neutron-noble gas scattering length can therefore be approximated, neglecting terms of $\mathcal{O}(<10^{-3}b_\text{N})$, as

\begin{equation}
b_\text{c}(q)^2 \approx b_\text{N}^2 + 2b_\text{N}b_\text{Y}(q) +2b_\text{N}b_\text{E}(q).
\label{eq:diff}
\end{equation}

\vspace{0.5cm}
If we assume the nucleus is infinitely-heavy and sufficiently isolated from neighboring nuclei, then effects from recoil, thermal motion, and interference arising from interatomic pair potentials are not present and the differential cross section and $b_\text{c}(q)^2$ are equivalent. We will assume this to be the case at the moment as it is sufficient for demonstrating the methodology of the measurement. We will later consider the small corrections more carefully in Sec.\,\ref{sec:sim}. 

\begin{figure}[!h]
\centering
\includegraphics[width = 0.5\textwidth]{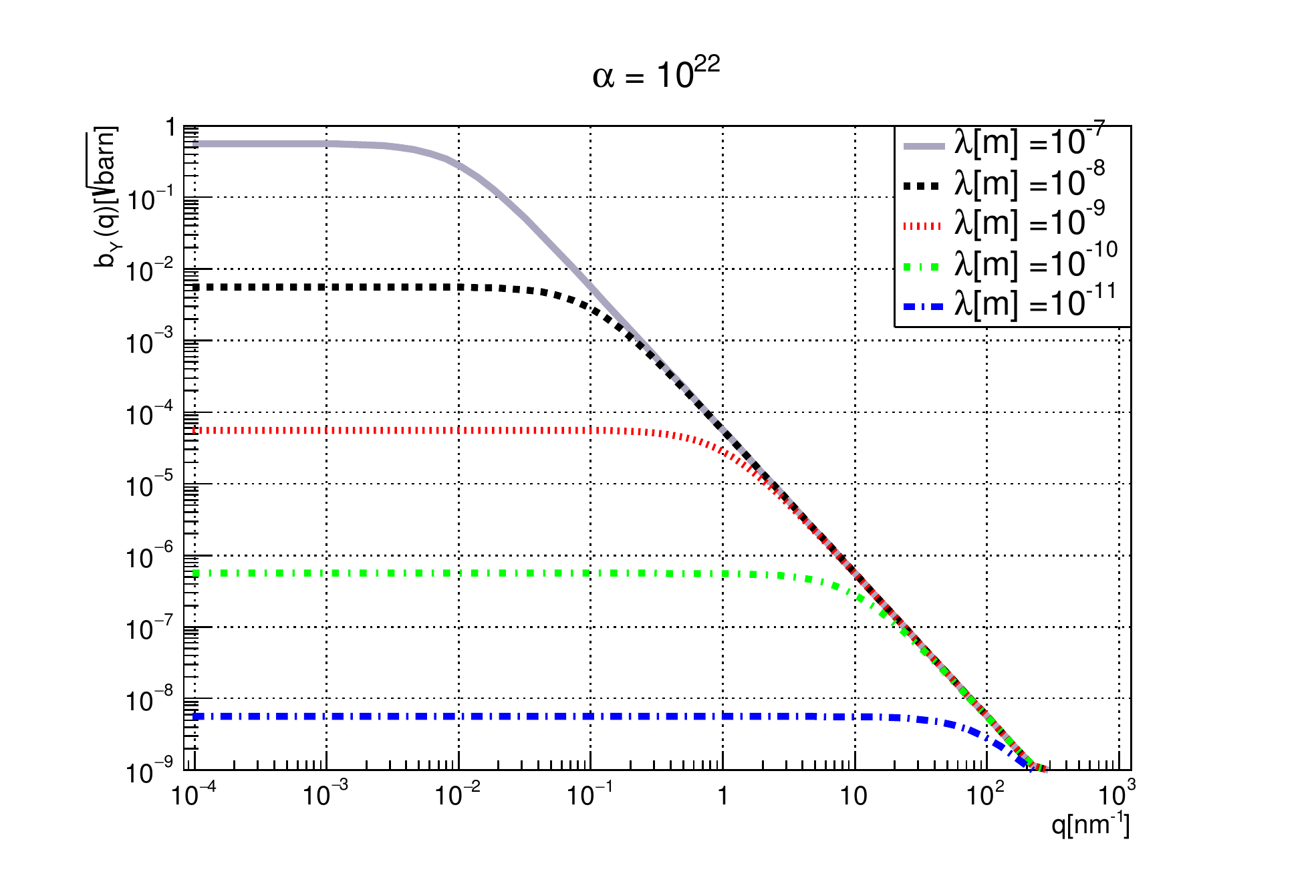}
\caption{Demonstration of the $\lambda$-dependence of $b_\text{Y}(q)$ for a fixed value of $\alpha$. Larger values of $q$ are sensitive to changes in $b_\text{Y}(q)$ for lower values of $\lambda$, and vice versa.}
\label{fig:byq}
\end{figure}

The $q$-dependence of the $b_\text{Y}(q)$ contribution to Eq.\,\ref{eq:diff} is drawn in Fig.\,\ref{fig:byq} for various values of interaction length $\lambda$.
There is a clear drop in intensity in the forward scattering region which is distinguishable from that of the nuclear potential known to be constant for meV energy neutrons in the absence of resonances. We can therefore search for ISL deviations at short range by measuring the $q$-dependence of neutron scattering intensity by free noble gas atoms. The noble gases are available in stable form at room temperature over a wide range of masses and their weak interatomic interactions minimize the size of the corrections to the scattering law from interference terms, further making them a suitable choice for this measurement.

\section{Experiment} %%%%%%%%%% Experimental setup %%%%%%%%%%%%

\subsection{Beamline  and Apparatus}

\begin{figure}[!ht]
\centering
\includegraphics[width = 0.45\textwidth]{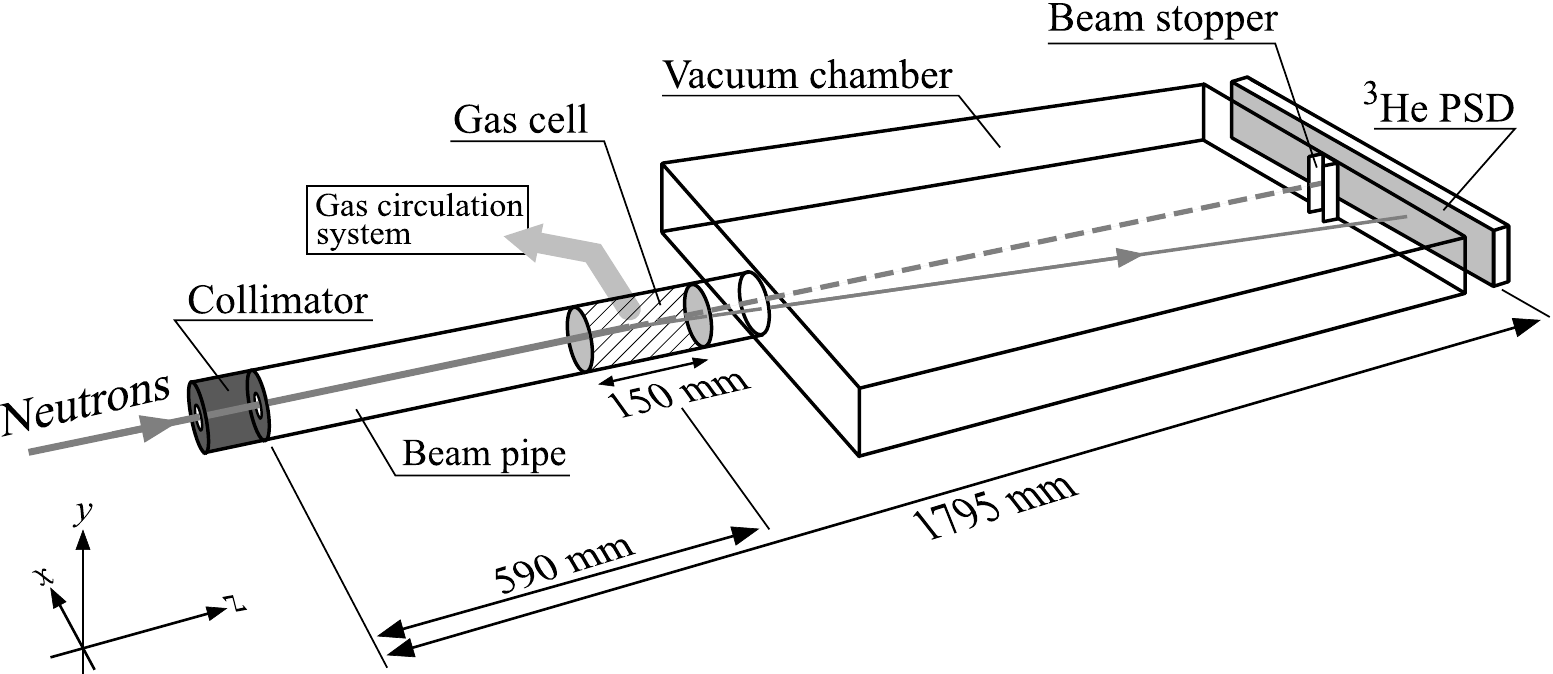}
\caption{Layout of our experiment as mounted on BL05.}
\label{fig:apparatus}
\end{figure}

The Neutron Optics and Physics (NOP) beamline is located at the Material Life Science Facility (MLF) at J-PARC and aims to study fundamental physics and neutron optics using pulsed-neutrons\,\citep{MLF,NN,design}. This experiment was performed on the Low-Divergence beam branch of NOP\,\citep{noriko}. A schematic view of our apparatus as mounted on NOP downstream of the beam duct is shown in Fig.\,\ref{fig:apparatus}. The coordinate system used in this paper is indicated in the figure: the $y$-axis is the upward vertical axis, the $z$-axis defines the direction of neutron beam travel, and the $x$-axis is chosen orthogonally so as to form a right-handed frame. 

%%%%%% Apparatus
%\input{320_apparatus}

Aluminum windows of thickness $0.1$\,mm divide the beam into three regions: the beam pipe, the gas cell and the vacuum chamber. The beam pipe was evacuated to a pressure of less than 3\,Pa up to $4.5\,$m upstream to avoid neutron scattering along the flight path. The gas cell was directly connected to the beam pipe with the aluminum windows on either end and 
was filled with the target gas. The center of the gas cell was located $590$\,mm downstream from a $10\,$mm(x) $\times$ $30\,$mm(y) beamline collimator. The cell had a diameter of $50$\,mm and a length of $150$\,mm giving it a volume of about $300$\,cm$^{3}$.
The inner surface of the gas cell was covered with a cadmium neutron absorbing layer. 

The downstream end of the gas cell was connected to a rectangular vacuum chamber allowing the scattered neutrons to propagate to the detector unobstructed and with sufficient angular resolution. 
The inner surfaces of the side walls of the scattering chamber were lined with several cadmium fins to absorb backscattered neutrons thereby reducing the number of background scattering events reaching the detector. 
The direct beam stopped at a beamstop on the downstream wall of the chamber. 
The beamstop was made of 30\,mm-deep, 90\,mm-tall box of 5\,mm-thick lithium fluoride plates to avoid contamination from scattered neutrons. 
The chamber had an exit window made of aluminum with a width of $600$\,mm, height of $100$\,mm, and thickness of $2$\,mm. 

Neutrons scattered by the target gas were detected by a $^3$He gas filled position 
sensitive detector (HePSD) upon exiting the scattering chamber. The HePSD contained seven proportional counter tubes each with a diameter of $12.7$\,mm altogether covering an area $600$\,mm wide and $90$\,mm high. Each tube had one-dimensional position sensitivity along the tube via charge division on a high resistivity anode. The horizontal position resolution was between 4 and 5\,mm for all tubes. The x-values of detected neutrons in each tube were calibrated using a pinhole-collimated neutron beam to scan the detector surface, and fitting the resulting mean detector-registered values of intensity as a function of collimator position. The relationship was found to be linear with an error of $<0.1\%$.
The vertical position resolution is defined by the tube diameter and is therefore $12.7$\,mm. Time-of-flight information was recorded to determine the incident neutron wavelength and momentum transfer. 
The tubes were filled with $^3$He gas with a partial pressure of $10^6\,$Pa and argon gas as a buffer. 
The applied voltage for the counter was 1530\,V. 
The detection efficiency was calculated to be $93.9\%$ for neutrons of wavelength of $0.3$\,nm.

\subsection{Gas handling}

If the target gas samples are contaminated with atmospheric gas molecules, $q$-dependence in the differential cross section will arise due to molecular diffraction\,\citep{outgas}. The gas samples introduced into our gas cell were of sufficiently high purity that this effect would be negligible, however molecules trapped in the gas cell walls may break free through a process called \enquote{outgassing}, and over time can result in significant systematic error. For example water vapor has a neutron cross section $\sigma_{H_2O}$ of about two orders of magnitude larger than the helium cross section $\sigma_{He}$ for neutrons with velocity $v\sim$1\,km/s. Therefore in order to measure the $q$-dependence of the scattering contribution from helium with a precision of $0.1\%$, competitive with existing data, the relative abundance of the trapped water molecules $\epsilon$ must satisfy the following relationship:

\begin{equation}
\epsilon\,\frac{\sigma_{\text{H}_2\text{O}}}{\sigma_{\text{He}}} < 10^{-3},
\end{equation}

\noindent giving a value of $\epsilon \approx 10\times 10^{-6}$, or 10\,ppm. Since helium has the lowest cross section of the noble gases and trapped $H_2O$ is the largest source of outgassing we estimate that 10\,ppm is the highest acceptable value of impurity contamination in the gas cell during measurement. 

In general one can \enquote{bake out} these trapped gases by heating up the gas cell while simultaneously evacuating, however the very thin aluminum windows of our gas cell would deform. Therefore we employ a continuous circulation and purification system to satisfy our purification requirements.

\begin{figure}[!ht]
\centering
\includegraphics[width = 0.45\textwidth]{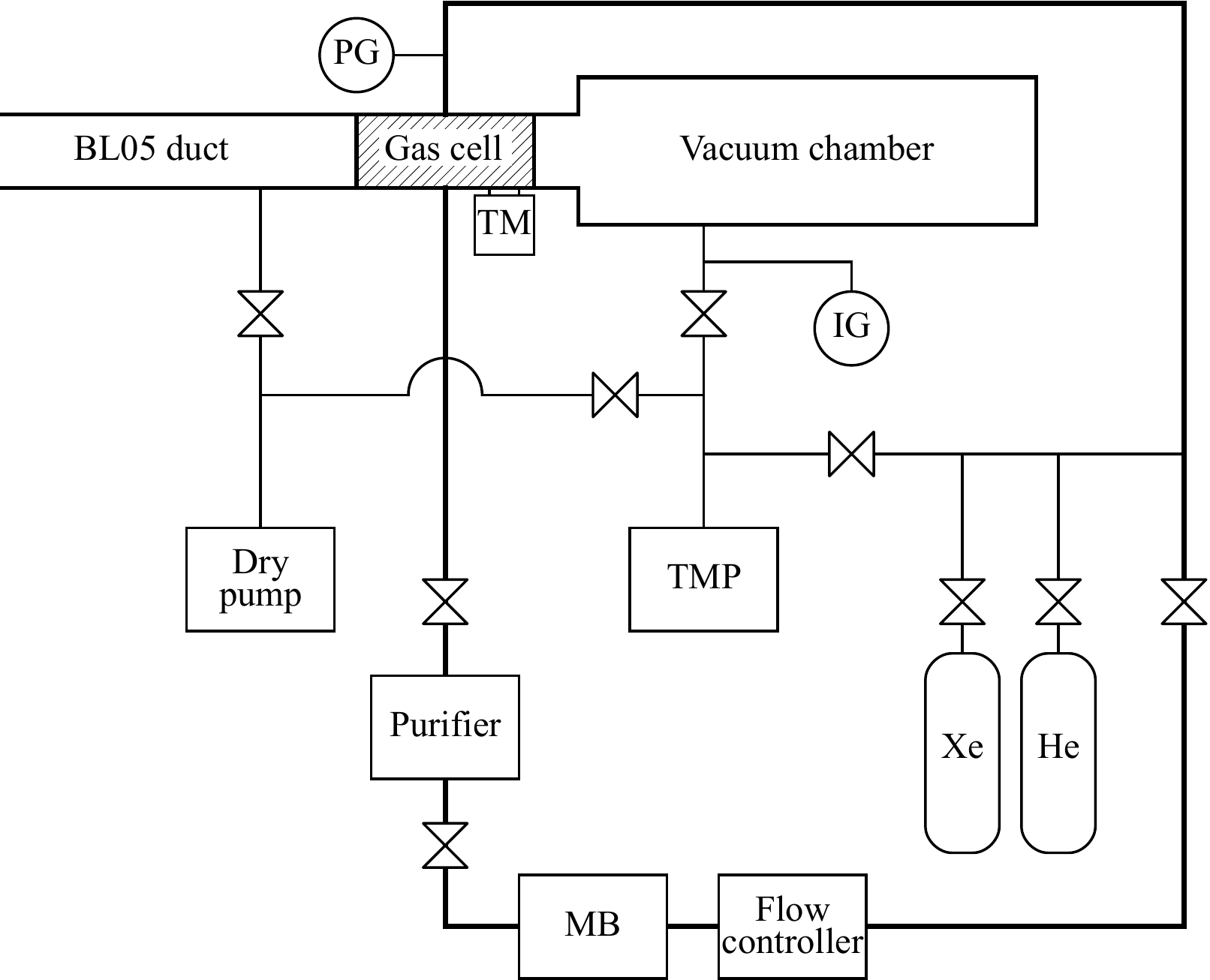}
\caption{Gas handling schematic. (TMP) turbomolecular pump,
   (MB) Metal Bellow pump, (PG) digital pressure gauge,
   (IG) ion gauge, (TM) platinum resistance thermometer.}
\label{fig:gas_schem}
\end{figure}

The gas handling and circulation system is outlined in Fig.\,\ref{fig:gas_schem}. The gas line consisted of 6.35\,mm diameter stainless steel tubing, Swagelok connections, and Swagelok valves type SS4H.
After evacuating the gas cell to less than $10^{-2}$\,Pa using a turbomolecular pump (TMP), the target gas was introduced to the system until a pressure of $190$\,kPa was achieved. A Metal Bellows IBS-151 pump (MB) and a flow controller provided gas circulation through a rare gas purifier SAES PS2-GC50, whose outlet gas purity for flow rates less than 200 sccm is rated to be less than $0.01$\,ppm for H$_{2}$O, N$_{2}$, O$_{2}$ or hydrocarbons, which is well below our stated upper limit of 10\,ppm, allowing plenty of room to improve upon current precision noble gas scattering measurements. 

The flow rate was controlled to 50\,cm$^{3}$/min, which means that the target gas was purified every 6\,min. This flow rate contributes an average bulk flow velocity of the target gas of $1.4\times10^{-4}$\,m/s, which is negligibly small compared to the velocities of  random thermal motion of the target gas atoms. The pressure and temperature of the gas cell were monitored during data acquisition with a digital pressure gauge Mensor CPG2400 (PG), an ion gauge Pfeiffer PKR251 (IG), and a platinum resistance thermometer PT100 (TM) with accuracies of 300\,Pa, 30\% at $<$100\,Pa,  and 60\,mK, respectively.

\subsection{Apparatus Characterization}

%\input{340_correcteddata}
%\subsubsection{Slit Scan}
\label{sec:slitscan}

\begin{figure}[!h]
\centering
\includegraphics[width=0.45\textwidth]{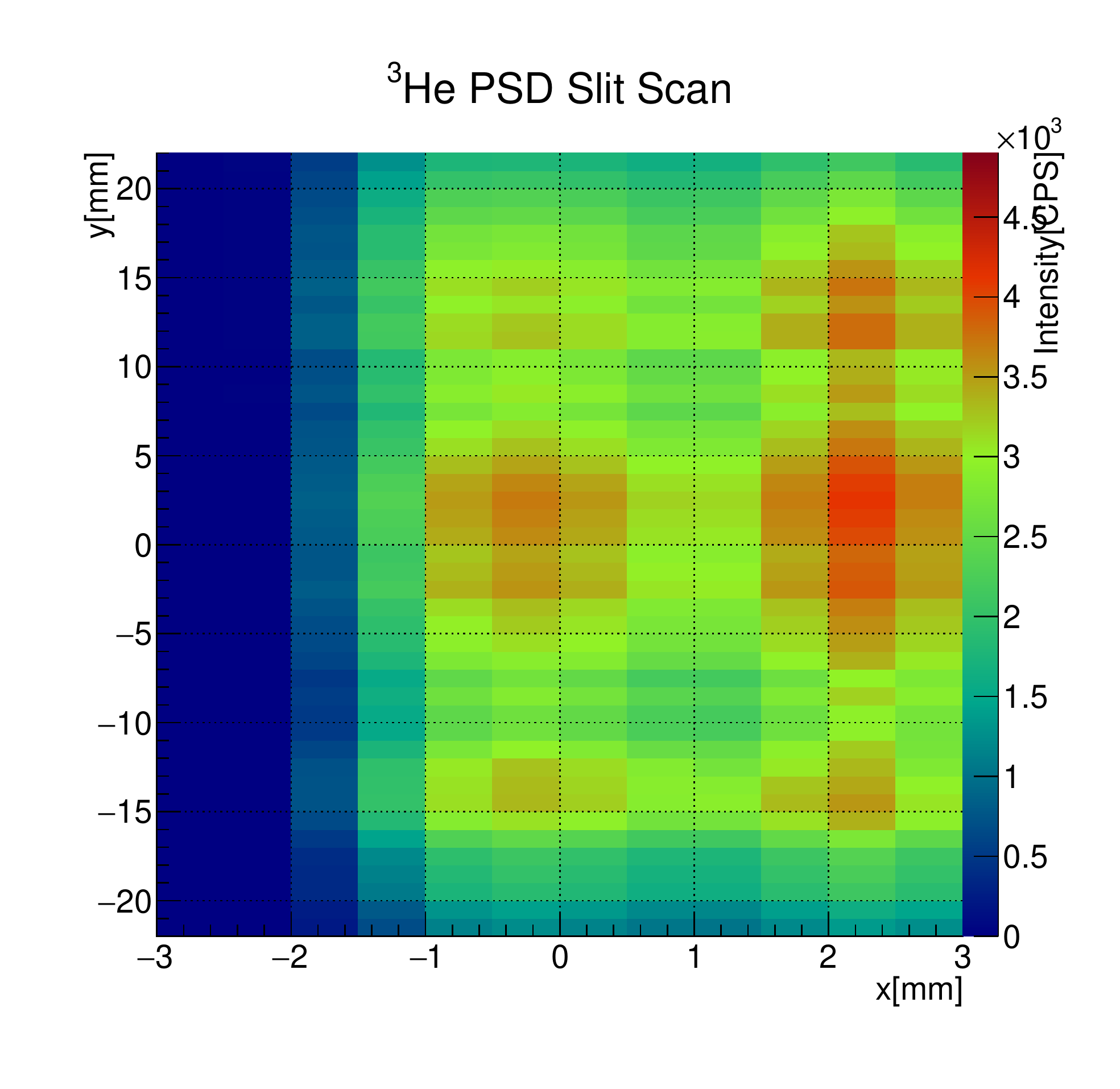}
\caption{2-D beam intensity profile during 150\,kW \\ proton beam power operation.}
\label{fig:slitscan}
\end{figure}

\begin{figure}[!h]
\centering
\includegraphics[width=0.45\textwidth]{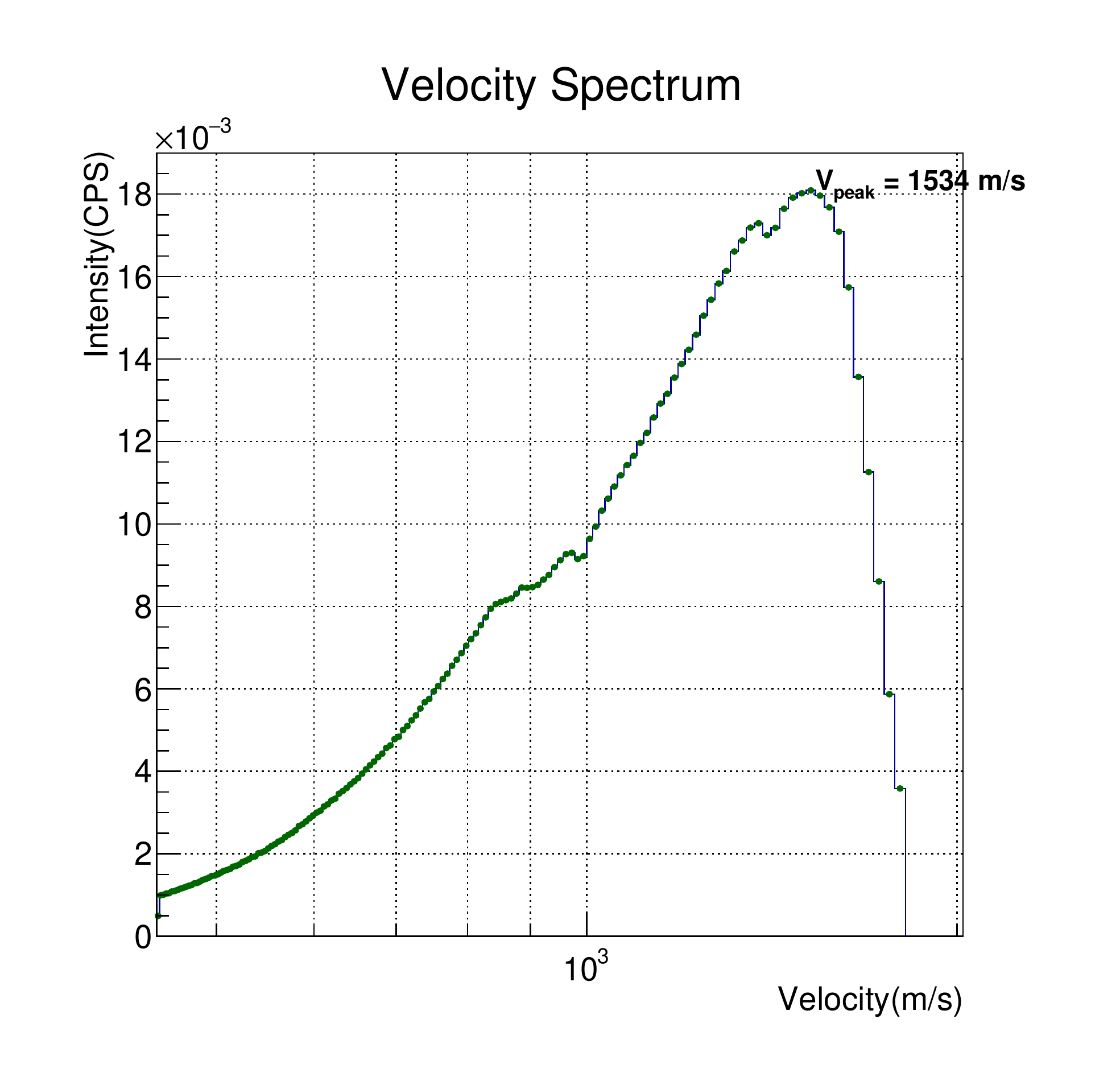}
\caption{Velocity distribution obtained from beam profile data.}
\label{fig:tofdist}
\end{figure}

We performed a $6$\,mm(x) $\times 44$\,mm(y) cross sectional scan of the beam using a pinhole collimator of diameter $\sim 1$\,mm, formed from two pairs of  B4C blades located $4.6\,$m upstream from the center of the gas cell. Data were recorded at each position and summed together to produce the plot shown in Fig.\,\ref{fig:slitscan}. This data also produced the velocity spectrum shown in Fig.\,\ref{fig:tofdist}.

\section{Pseudoexperimental Simulation} %%%%%%%%%%%%%%%% Simulation %%
\label{sec:sim}
%%%%%%%%%%%%%%%%%

%\input{400_simulation}
In order to uncover the single neutron-noble gas atom differential cross section from the total measured gas scattering intensity spectrum, it is crucial to remove effects arising from the finite beam cross section and gas cell size, incident beam intensity nonuniformity, beam divergence, and distribution of incident neutron energies. This is done by first performing a pseudoexperimental simulation which will be used to normalize the experimental data in order to look for residual $q$-dependences which can be used to place a limit on the strength of a possible Yukawa-like interaction.
  
The experiment was simulated using the Monte Carlo method implemented within the ROOT analysis framework. Neutrons are generated in a loop and propagated to and from scattering centers using standard kinematic relations. The initial neutron phase space was determined using the measured time of flight and two-dimensional scans of neutron intensity on the beam line described in Sec.\,\ref{sec:slitscan}.

The probability of a neutron scattering from a gas atom is computed using the well-known function for neutron transport through uniform density macroscopic materials,

\begin{equation}
P = 1 - e^{-\rho\,z\,\sigma_\text{s}}\,
\end{equation}

\noindent where $\rho$ is the number density of the scatterers, $z$ is the position of the neutron within the material along the beam axis, and $\sigma_\text{s}$ is the energy and temperature dependent scattering cross section for a free nucleus. For a neutron of energy $E_\text{n}$ and a gas of mass number $A$ at temperature $T$,

\begin{equation}
\sigma_\text{s} = \frac{\sigma_\text{c}}{(1+\frac{1}{A})^2}\bigg(1+\frac{1}{2\lambda}\bigg)\text{Erf}\bigg\lbrack\sqrt{\lambda}+\frac{e^{-\lambda}}{\pi\lambda}\bigg\rbrack
\end{equation}

\noindent where $\lambda = \frac{E_\text{n}\,A}{k_\text{B}T}$, and $k_\text{B}$ is the Boltzmann constant.

Upon scattering, the final energy $E_\text{n}^\prime$ and scattering angle are chosen according to the double differential cross section 

\begin{equation}
\frac{d^2\sigma_\text{s}}{d\Omega\,dE_\text{n}^\prime} = b_\text{c}^2(q)\,S(q,\omega)
\end{equation}

\noindent where $b_\text{c}$ is the coherent scattering length and $S(q,\omega)$ is the so-called dynamic structure factor which describes the response of the scattering system in the ($q,\omega$) plane where $q$ and $\omega$ are the momentum and energy transfer to the system, respectively. Explicitly $q$ and $\omega$ are given in terms of the incoming neutron energy $E$, outgoing neutron energy $E^{\prime}$, and scattering angle $\theta$ as

\begin{align}
q &= \sqrt{2m/\hbar^2}\sqrt{E+E^{\prime}-2\sqrt{EE^{\prime}}\cos(\theta)},\label{eq:qreal}\\
\omega &= \frac{E-E^{\prime}}{\hbar}.
\end{align}

The function $S(q,\omega)$ is known analytically for the case of an ideal gas of atomic mass $M$ in thermal equilibrium at temperature $T$ which follows a Boltzmann velocity distribution and is given by\,\citep{lovesey}

\begin{equation}
S(q,\omega) = \frac{\beta M}{2\pi\hbar^2q^2}\text{exp}\bigg\lbrack
-\frac{\beta M}{2\hbar^2q^2}\bigg(
\hbar\omega-\frac{\hbar^2q^2}{2M}
\bigg)^2
\bigg\rbrack,
\label{eq:sqw}
\end{equation}
\vspace{3mm}
 
\noindent where $\beta \equiv 1/k_BT$. Note that this is one of the few cases in which one can present an exact expression for $S(q,\omega)$ from first principles. The ability to use this known functional form to fit the scattering data and search for possible exotic interactions is a key feature of our method. Deviations from Eq.\,\ref{eq:sqw} due to the finite size of the gas atoms and the long range attractive interatomic interactions are negligible for pressures $\sim\mathcal{O}$(1\,atm) and values of $q$ such that $qR\ll\,1$, where $R$ is the radius of the gas atom. They are also in principle calculable to high accuracy. In fact these calculations are nothing more than the well-known Mayer cluster expansions discussed in statistical mechanics textbooks, which are the best-understood many body calculation in physics after more than a century of theoretical effort. These corrections can be related in principle to the measurements of the virial coefficients of the noble gases. Our method is not limited in precision by uncertainties in $S(q,\omega)$ and can therefore be improved in future work.

Once scattered, the neutron is propagated to the detector according to the experimental geometric constraints. The ROOT framework then allows us to store the final phase space coordinates to be viewed and compared with experimental data. However our experimental setup does not distinguish incident energy $E$ from outgoing energy $E^{\prime}$ but measures the total time of flight from moderator to detector to compute the neutron energy. We therefore parametrize our results in terms of $q$ in the absence of inelastic scattering, $q\rightarrow \sqrt{\frac{2mE}{\hbar^2}}\sin(\theta/2)$ and find through simulation any possible loss in accuracy to the precise value of $\alpha$ using this parametrization method.

%By running the simulation code with different values of $\alpha$ we take the ratio $I_{\alpha}(q)/I_0(q)$, where $I_{\alpha}(q)$ and $I_0(q)$ are the detected neutron intensity distributions for the case of a nonzero and zero Yukawa term, respectively, and fit the result to a function of the form $a + b\times f_\text{Y}(q)$. We then compare the fit results to the input values of $\alpha$. We found we could recover the input value of $\alpha$ to within $20\%$, implying an $\alpha$ resolution of $80\%$, which is sufficient in placing an upper limit on its value so long as we adjust our experimental fit results accordingly.

\section{Analysis} %%%%%%%%%%%%%%%%%% Analysis %%%%%%%%%%%%%%%%%%%
\label{sec:ana}

%\input{500_analysis}
% This is a General procedures in Analysis.
% general procedure
% direct beam correction, solid angle correction
% vacuum data subtraction

To isolate the neutron-gas scattering, $I^{\text{gas}}(q)$, from background we subtract from the total intensity the vacuum cell condition $I^{\text{vac}}(q)$. This removes scattering effects from the aluminum gas cell windows and downstream scattering chamber window, as well as the effect of neutron-beamstop scattering. Before subtraction, the vacuum cell intensity is corrected to account for the TOF-dependent neutron absorption cross section of each gas.   
% take the ratio of He data

To further isolate the $q$-dependence of the Yukawa-like interference term we normalize the experimental $I^{\text{gas}}(q)$ distribution by the corresponding distribution generated using the pseudo-experimental simulation technique described in the previous section without introducing a Yukawa term to the interaction. We first express the experimental and simulated distributions as 

\begin{align}
I_{\text{EXP}}(q) &= C_{\text{EXP}}(\Omega)\times\left(b_\text{i}^2+ b_\text{c}^2+2\,b_\text{c}b_{\text{E}}(q)+2\,b_\text{c}\,b_{\text{Y}}(q)\right)\nonumber\\
I_{\text{SIM}}(q)& = C_{\text{SIM}}(\Omega)\times\left( b_\text{c}^2+2\, b_\text{c}b_{\text{E}}(q)\right)
\label{eq:Iq}
\end{align}
\vspace{1mm}

\noindent where $C_{\text{EXP}}(\Omega)$ and $C_{\text{SIM}}(\Omega)$ are solid angle ($\Omega$)-dependent functions which relate $I(q)$ to the differential cross section. In principle they are equivalent however due uncertainty in the interaction point within the cell (assumed to be center of the cell when constructing $q$ from the experimental data), finite detector resolution, and beam divergence they may differ in a gas species-independent way by a few percent. In order to remove this effect we form the ratio, neglecting terms of $\mathcal{O}(<10^{-3}\,b_\text{c}$): 

\begin{equation}
R(q) \approx \frac{C_{\text{EXP}}(\Omega)}{C_{\text{SIM}}(\Omega)}\times \bigg(\frac{b_\text{i}^2+ b_\text{c}^2}{b_\text{c}^2}+2\frac{b_\text{Y}(q)}{b_\text{c}}\bigg).
\end{equation}
\vspace{1mm}

\noindent where $b_i$ is the incoherent scattering length. Note that $b_i$ was not included in the simulated distribution. This was due to the very limited amount of published data. Fortunately this term only contributes an isotropic background which we may account for by using an additional fit parameter. 

The function $R(q)$ is found for two gas species $j, k$ and the respective ratios $R^{j}(q)$ and $R^{k}(q)$ are obtained. We then form an additional ratio $R^{jk}(q)\equiv R^{j}(q)/R^{k}(q)$ which removes the $\Omega$-dependence at the cost of reducing our sensitivity to $b_\text{Y}(q)$,

\begin{equation}
R^{jk}(q) \approx \bigg(P_0+2\bigg\lbrack\frac{b_{\text{Y}}^j(q)}{b_{\text{c}}^j}-\frac{b_{\text{Y}}^k(q)}{b_{\text{c}}^k} \bigg\rbrack\bigg),
\label{eq:super}
\end{equation}
\vspace{1mm}

\noindent where constant terms that include incoherent scattering constributions are absorbed in $P_0$, and we again neglect terms of $\mathcal{O}(<10^{-3}\,b_\text{c})$. By choosing gases which are very different in mass, the loss in sensitivity to $\alpha$ due to the difference term in Eq.\,\ref{eq:super} can be minimized. For the case of Xe ($A = 131$, $b_\text{c} = 4.69(4)$\,fm) and He ($A = 4, b_\text{c} = 3.26(3)$\,fm)\,\citep{Rauch}, the loss in sensitivity alpha is only $1\%$. Therefore one can produce an exclusion plot in the $\alpha-\lambda$ plane by first normalizing the experimental $I(q)$ data for two separate gases by the respective simulated $I(q)$ distributions, then performing a simple $\chi$-squared minimization with Eq.\,\ref{eq:super} to the ratio of these results using various values of $\alpha$ and $\lambda$.

\section{Results and Discussion} %%%%%%%%%%%%%%% Results and Discussion %%%%%%%%%%%%%
\subsection{Summary of Data Analysis}
During our run on BL05 the beam power to MLF was an average of $ 
150$\,kW (1.1$\times 10^6$ incident neutrons/sec) and we recorded $2.5\times 10^6$ scattering events for evacuated cell, $2.8\times 10^6$ for He gas-filled cell, and $1.1\times 10^7$ scattering events for the Xe gas-filled condition. Each event was recorded with time of flight and position information which allowed us to construct the scattered intensity $I(q)$ using the elastic scattering parameter (see end of Sec.\,\ref{sec:sim}) 

\begin{equation}
q = \frac{4\pi}{\lambda_n}\sin\bigg(\frac{\theta}{2}\bigg)\,
\label{eq:qel}
\end{equation}

\noindent where $\lambda_n$ is the neutron wavelength before scattering, and $\theta$ is the scattering angle as measured from the center of the gas cell to the center of the detector. Additional information including gas cell temperature and pressure, and the proton beam power were recorded and used to normalize the data.

\begin{figure}[!h]
\centering
  \centering
  \includegraphics[width=0.5\textwidth]{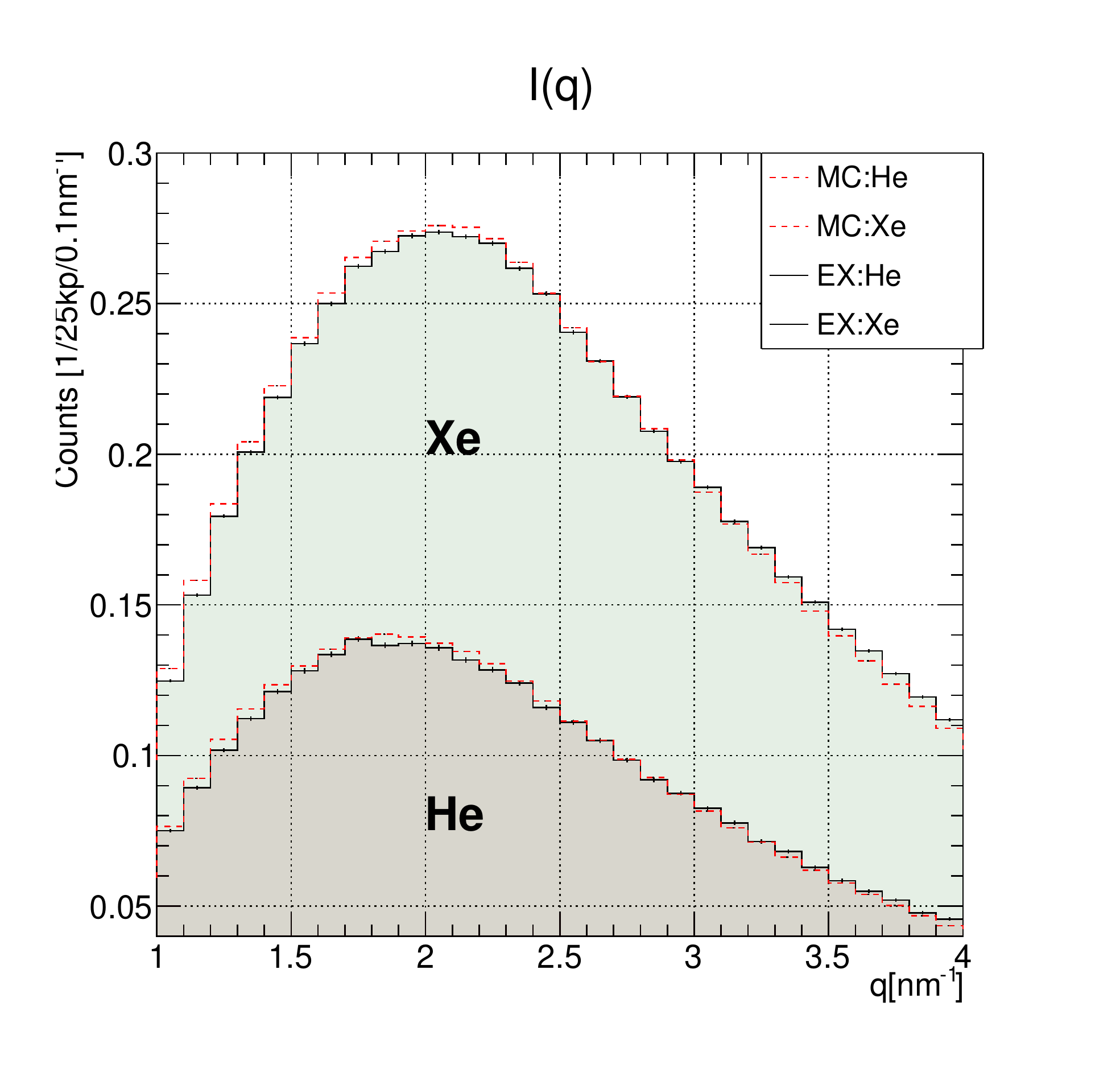}
  \caption{Raw $I(q)$ distribution compared with Monte Carlo simulation results.}
  \label{fig:EXMCcompare}
\end{figure}

\begin{figure}[!h]
  \centering
  \includegraphics[width=0.5\textwidth]{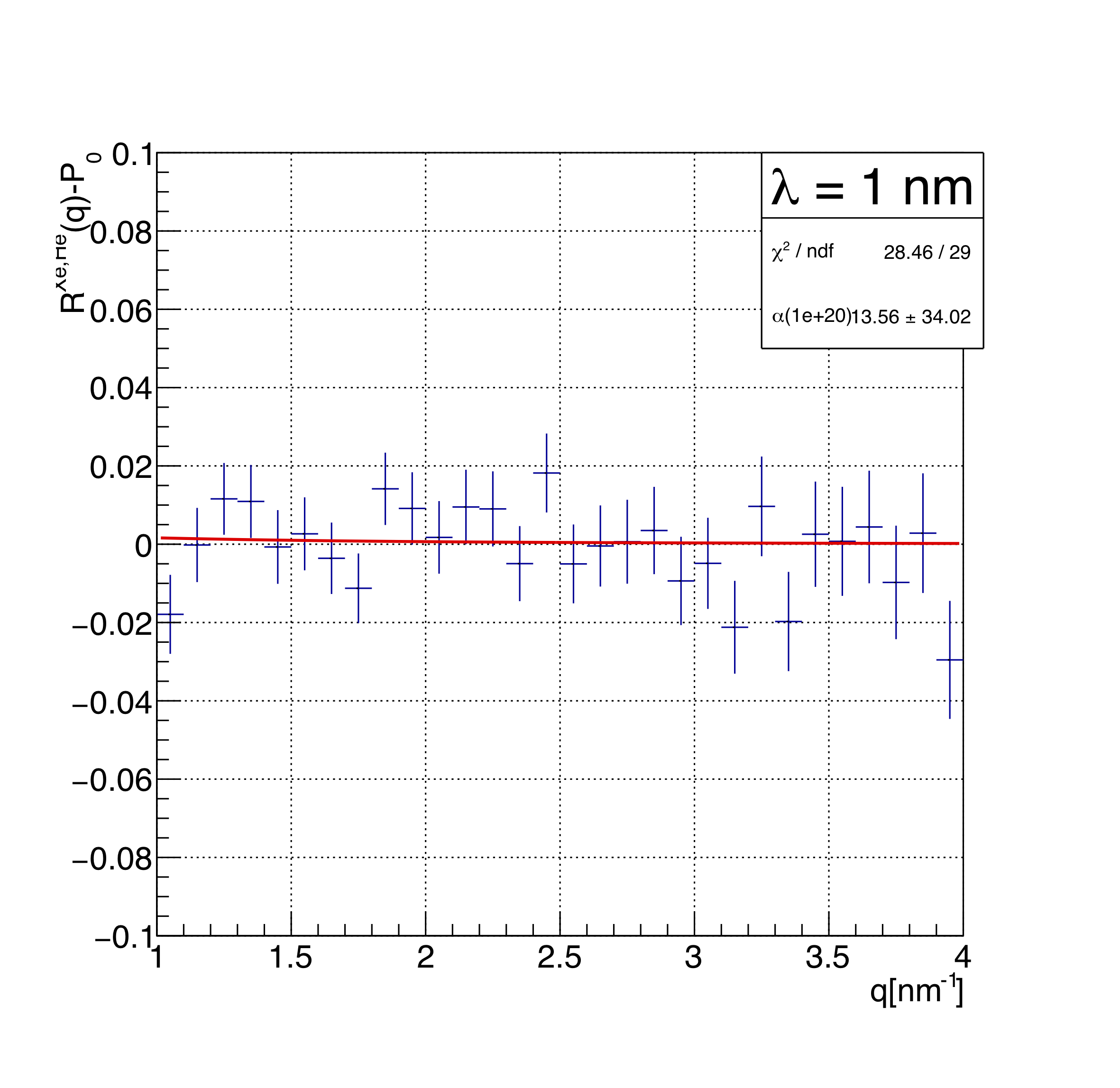}
  \caption{Residual $I(q)$ distribution used to extract the value of $\alpha$ as a function of $\lambda$.}
  \label{fig:res}
\end{figure}

A direct comparison of the measured $I(q)$ distributions to the simulated distributions are shown in Fig.\,\ref{fig:EXMCcompare}. Following the procedure outlined in Sec.\,\ref{sec:ana}, we form a ratio of the simulation-normalized data minus the constant contributions from incoherent scattering in Fig.\,\ref{fig:res}. 

Before fitting our data with an analytic function of $q$ it was important that we accounted for the fact that the elastic parametrization of $q$ used for data acquisition, Eq.\,\ref{eq:qel}, is different from the true value given in Eq.\,\ref{eq:qreal}. To account for this we first ran the simulation code with several different values of $\alpha$. We then took the ratio $I_{\alpha}(q)/I_0(q)$, where $I_{\alpha}(q)$ and $I_0(q)$ are the detected neutron intensity distributions for the case of a nonzero and zero Yukawa term, respectively, and fitted the result to a function of the form $a + b\times f_\text{Y}(q)$. We then compared the fit results to the input values of $\alpha$ and found we could recover the input value of $\alpha$ to within $20\%$, implying an $\alpha$ resolution of $80\%$, which is sufficient in placing an upper limit on its value so long as we adjust our experimental fit results accordingly.

We therefore use the following fit function of the form Eq.\,\ref{eq:super} adjusted for the $80\%$ reduction in sensitivity due to $q$ parametrization, 

\begin{equation}
R^{\text{fit}}(q) = 0.8\times\,2\bigg\lbrack\frac{b_{\text{Y}}^j(q)}{b_{\text{c}}^j}-\frac{b_{\text{Y}}^k(q)}{b_{\text{c}}^k} \bigg\rbrack,
\end{equation}

\noindent to find the least squares best fit value for $\alpha$ as a function of $\lambda$. Shown in Fig.\,\ref{fig:res} is the result for the case of $\lambda = 1\,$nm.


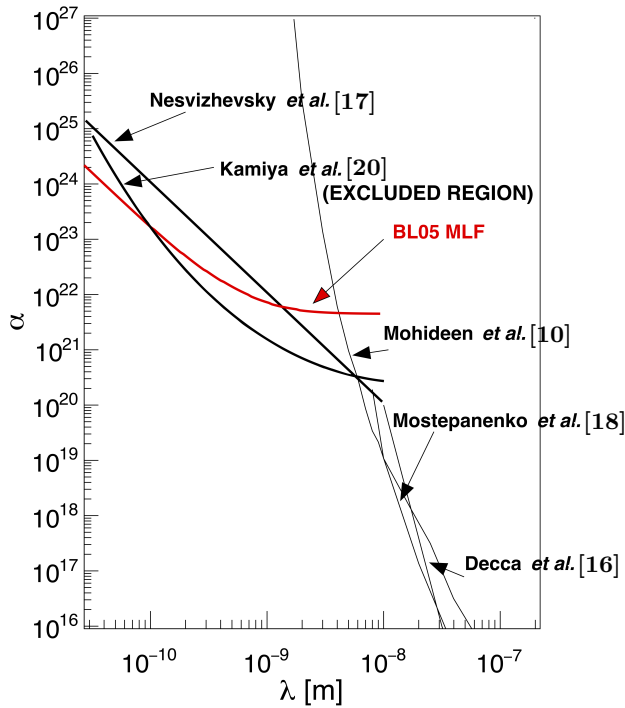
\begin{figure}
\begin{tikzpicture}
    \node[anchor=south west,inner sep=0] at (0,0) {\includegraphics[width = 0.45\textwidth]{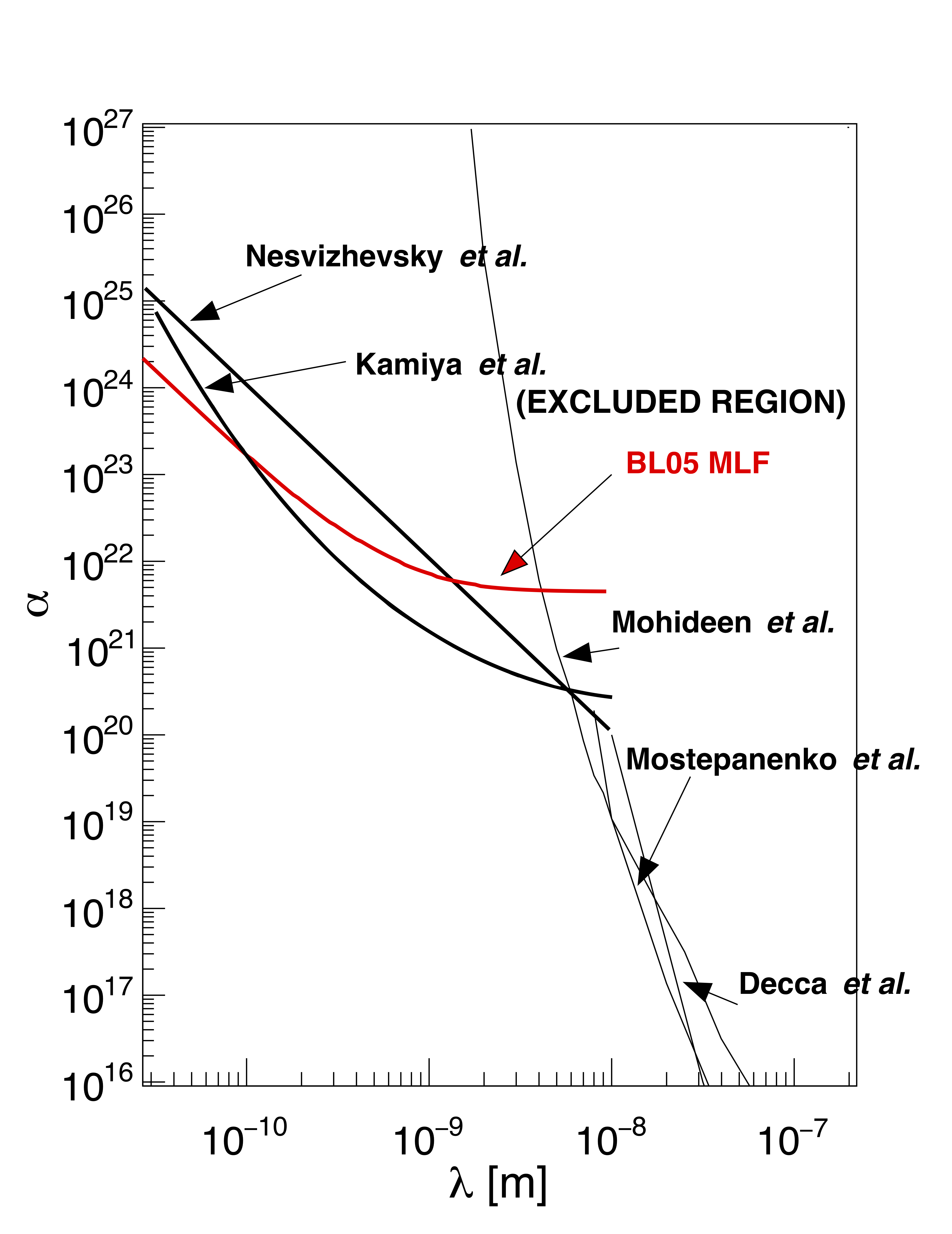}};
    \filldraw[white] (4.4,8.25) circle (0.1pt) node[anchor=west] {\bf\color{black}\cite{Nez08}};
    \filldraw[white] (4.56,7.37) circle (0.1pt) node[anchor=west] {\bf\color{black}\cite{kamiya}};
    \filldraw[white] (7.0,5.16) circle (0.1pt) node[anchor=west] {\bf\color{black}\cite{Mohideen98}};
    \filldraw[white] (7.73,4.0) circle (0.1pt) node[anchor=west] {\bf\color{black}\cite{mp2008}};
    \filldraw[white] (7.65,2.1) circle (0.1pt) node[anchor=west] {\bf\color{black}\cite{Decca}};    
\end{tikzpicture}
\caption{Exclusion curve in the $\alpha-\lambda$ plane resulting from our analysis, labeled \enquote{BL05 MLF} and superimposed over results of several other experiments.}
\label{fig:el}
\end{figure}

This method was repeated for $100$ different values between $\lambda = 0.01$\,nm and $\lambda = 10$\,nm. The lower bound was chosen so as to remain consistent with our assumptions in Sec.\,\ref{sec:method}, namely that effects due to $b_\text{M}(q)$ arising from interactions between the neutron's magnetic dipole moment and the slowly varying electric and magnetic fields of the scattering centers remains negligible. The upper bound was chosen simply because the sensitivity of our method is greatly overshadowed by former experiments described in Sec.\,\ref{sec:intro}. All values of $\alpha$ as a function of $\lambda$, $\alpha(\lambda)$, obtained had a central value which was consistent with zero within one standard deviation.\\
\iffalse
\begin{figure}[!h]
\centering
  \centering
  \includegraphics[width=0.5\textwidth]{raw_exp_limit.pdf}
  \caption{Experimentally obtained values of $\alpha(\lambda)$ drawn with a 95\% confidence ($1.96\,\sigma$) exclusion band.}
  \label{fig:raw_limit}
\end{figure}
\fi
\indent Under the assumption that each measured value of $\alpha$ follows a Gaussian distribution about the true mean, we produced an exclusion plot in the $\alpha-\lambda$ plane by choosing the upper bound of a symmetric interval about $\alpha = 0$ while preserving the area under the Gaussian for each value of $\alpha(\lambda)$. For example when $\lambda = 0.1$\,nm, we find that the true value of $\alpha$ lies within the interval $\alpha(0.1\,\text{nm}) = (1.6\pm 16.2)\times 10^{22}$ with 95\% confidence. When expressed as a symmetric interval about zero with equal confidence the interval becomes $\alpha(0.1\,\text{nm}) = (0.0\pm 16.5)\times 10^{22}$. By reporting as a symmetric interval about zero we exploit the fact that our measurement is sensitive to both positive and negative values of $\alpha$. The upper bounds of $\alpha(\lambda)$ are shown in Fig.\,\ref{fig:el}, superimposed over results of several other experiments sensitive to $\alpha$ in this $\lambda $ region.\\
A similar neutron scattering experiment was recently performed by Kamiya {\it et al.} at the HANARO research reactor of the Korean Atomic Energy Research Institute\,\citep{kamiya} which was also used to constrain $\alpha$ over a similar range of $\lambda$ range and is also shown in Fig.\,\ref{fig:el}. In that experiment the $q$ acceptance region was about one order of magnitude lower than ours and was therefore more sensitive in the larger $\lambda$ region (see Fig.\,\ref{fig:byq}).

\subsection{Systematic Effects}

%Our experiment was performed on a pulsed neutron beamline and is therefore capable in principle of a much larger $q$ acceptance region than what we report in this paper. We were limited in this experiment by some technical difficulties, which will be corrected during future data runs.  

Since the majority of incident neutrons do not scatter in the gas cell region ($>99\%$) it is crucial that the beamstop is constructed and installed properly to reduce transmitted neutrons and prevent neutron-beamstop backscattered neutrons from reentering the detector region.

During analysis of the direct beam data a left-right asymmetry about the direct beam axis for small scattering angles was apparent indicating that our beamstop was not properly centered and may have been slightly rotated about the vertical axis. This forced us to increase the lower bound of our angular acceptance as we had to cut a larger region from the central detector region ($\pm 100$\,mm from center) than originally anticipated ($\pm 50$\,mm from center) to remove this unwanted data. This restricted our usable $q$ region to $q>1\,$nm$^{-1}$ (originally we anticipated a lower bound of $q=0.5$\,nm$^{-1}$). This effect can be easily reduced in the future by simply employing a more careful beam-beamstop alignment during apparatus characterization and adding an additional thin layer of neutron absorbing material, such as cadmium, to all surfaces of the beamstop to further prevent unwanted neutron transmission. We estimate that upon proper beam-beamstop alignment and the addition of a $1$\,mm cadmium layer to all surfaces we can reduce the lower bound of $q$ by at least a factor of two.

In the larger $q$ region we found that the measured neutron intensity distribution dropped off more quickly than expected for isotropic scattering in the q region beginning at $q\approx\,3$\,nm$^{-1}$ for both He and Xe. This was in fact predicted by simulation through implementation of the finite size of the gas cell which indicated that neutrons which were scattered within the gas cell at sufficiently large angles were blocked from reaching the detector by absorption in the inner side walls of the cell aligned with cadmium. However the drop off was not as sharp as we had expected and differed significantly for values of $q>4$\,nm$^{-1}$. Due to the disagreement in this region we decided not to include that data in the final cut, originally intended to extend to $q=6$\,nm$^{-1}$. 

We have identified two possible sources for the origin of the excess counts in the larger $q$ region. First, Neutrons which undergo multiple scattering from the inner walls of the vacuum chamber following backscattering from the beamstop may enter the detection region at larger angles (See Fig.\,\ref{fig:apparatus}), mimicking gas scattering events. In this case we could extend the side walls of the beam catcher along the direction of the neutron beam, forming a rectangular tube extending backwards into the vacuum chamber, which would greatly reduce the solid angle of neutrons backscattered from the beam catcher. The second possible source could be due to the deformation of the 0.2\,mm thick aluminum gas cell windows resulting from the $\sim 2$\,atm pressure differential. This effect can be studied in future data runs where we will vary the gas pressure over several atmospheres and study any resulting change in the intensity distribution in this region. If this is found to be the source of the excess count rates we can optimize the thickness of the windows so as to minimize the deformation effect while maximizing neutron transmission.    

\section{Conclusion} %%%%%%%%%%%%%%%%%%%%%%% Conclusion %%%%%%%%%%%%%%%%%%%%%%%%%%

We performed a neutron-noble gas scattering measurement on BL05 at the MLF facility at J-PARC which places a limit on the upper bound of the strength of a possible deviation from the inverse square law of gravity of Yukawa form in the $\sim\,$nm range with $95\%$ confidence. We improved upon the limit placed by a recent similar experiment performed by Kamiya \textit{et al.} in the region below $\lambda = 0.1\,$nm due to the higher upper limit of the $q$ acceptance of our apparatus.

Unanticipated systematic errors were discernible near the lower and upper regions of our $q$ acceptance, thereby reducing the data available for gas scattering analysis, and thus our statistical sensitivity suffered. We plan to recover the full range of $q$ initially sought ($q=0.5$\,nm$^{-1}$ to $q=6$\,nm$^{-1}$) by adding an additional thin layer of neutron absorbing material to our beamstop and implementing a proper beamstop alignment procedure, which we believe will reduce the lower bound of the $q$ acceptance by a factor of two. To extend the upper bound of $q$ we seek to experimentally identify the possible sources of the systematic error in count rate we encountered in the larger $q$ region by making changes to the geometry of our beam stopper, as well as performing a study of the effect of gas pressure in the cell as we vary it over several atmospheres.

Although it is not possible to increase the flight path on the current beam line due to space restrictions, it is possible to increase the detection resolution by implementing a scintillation based PSD with 1\,mm resolution, as opposed to our current PSD which has 4-5\,mm resolution. Combined with focusing neutron optics to reduce the width of the beam without sacrificing statistical sensitivity, we can imagine increasing our sensitivity in the low $q$ region, so that with a comparable amount of beam time on BL05 we can repeat our measurement and improve on the existing limits on $\alpha$ in the region above $\lambda = 1\,$nm.

We also plan to analyze our data to constrain other types of possible exotic interactions in this regime with different functional forms.

\section*{Acknowledgments} %%%%%%%%%%%%%%%%%%% Acknowledgments %%%%%%%%%%%%%%%%%%%%%%%%

This work was supported by MEXT KAKENHI grant number JP19GS0210 and JSPS KAKENHI grant number JP25800152. We wish to thank the help given by Setsuo Sato for detector and software operation.
Work at the facility of J-PARC was performed under an S-type project of KEK (Proposal No. 2014S03) and user programs (Proposal No. 2016B0212, 2016A0078, and 2015A0239). C. Haddock acknowledges support from the Japan Society for the Promotion of Science. C. Haddock and W. M. Snow acknowledge support from NSF grant PHY-1614545 and from the Indiana University Center for Spacetime Symmetries.

\bibliographystyle{plain}

\end{document}